\documentstyle[aps,multicol,fleqn,epsf,psfig]{revtex}

\newcommand{\beq}{\begin{equation}}
\newcommand{\eeq}{\end{equation}}
\newcommand{\beqa}{\begin{eqnarray}}
\newcommand{\eeqa}{\end{eqnarray}}
\newcommand{\ket} [1] {\vert #1 \rangle}
\newcommand{\bra} [1] {\langle #1 \vert}

\begin{document}

\title{Security of quantum key distribution using $d$-level systems}

\author{Nicolas J. Cerf,$^{1,2}$ Mohamed Bourennane,$^3$ Anders Karlsson,$^3$
and Nicolas Gisin$^4$}

\address{$^1$ Ecole Polytechnique, CP 165, 
Universit\'e Libre de Bruxelles, 1050 Bruxelles, Belgium\\
$^2$ Jet Propulsion Laboratory, California Institute of Technology, 
Pasadena, California 91109\\
$^3$ Department of Microelectronics and Information Technology,
Royal Institute of Technology (KTH),\\ Electrum 229, SE-164 40 Kista, Sweden\\
$^4$ GAP-Optique, Universit\'e de Gen\`eve, 20 rue de l'Ecole de M\'edecine, 
Gen\`eve 4, Switzerland}

\date{July 2001}
\draft

\maketitle

\begin{abstract}
We consider two quantum cryptographic schemes relying on encoding
the key into {\em qudits}, i.e. quantum states in a $d$-dimensional 
Hilbert space. The first cryptosystem uses two mutually unbiased bases
(thereby extending the BB84 scheme), while the second exploits
all the $d+1$ available such bases (extending the six-state
protocol for qubits). We derive the information gained by
a potential eavesdropper applying a cloning-based individual attack, 
along with an upper bound on the error rate that ensures 
unconditional security against coherent attacks.
\end{abstract}


\begin{multicols}{2}

Quantum key distribution is probably one of the most promising concepts
in quantum information theory, and has been extensively studied
both theoretically and experimentally since its discovery by Bennett
and Brassard in 1984\cite{BB84}. 
This cryptographic method allows two remote parties
to share a secret key by use of a quantum channel supplemented with 
a public authenticated classical channel (see e.g. \cite{gisin} for 
a review). The impossibility for an eavesdropper to tap the quantum channel
without disturbing the communicated quantum data -- in a way that can,
in principle, be detected using the classical channel -- 
ensures the security of the key distribution. 
Most of the research effort to date has focused on quantum cryptosystems
based on two-dimensional quantum variables (qubits) carried e.g. by
the polarization state of individual photons. In particular, the optimal
individual attack is now known both for the BB84 protocol (using two 
mutually unbiased bases\cite{Fuchs}) 
and for the six-state protocol (using all three 
maximally unbiased bases\cite{DBruss,HBech-NGisin}). 
Strong bounds have also been derived in the
more general case of coherent attacks, which are useful to assess
the security of quantum cryptography (see e.g. \cite{Mayers,Lo,Shor}).
For higher-dimensional systems, however, very few results have been
obtained on the resistance to eavesdropping of {\em qudit}-based schemes 
(i.e., schemes based on encoding the key on $d$-level systems).
The only schemes that have been
considered use either two bases for a ququat (4-level system) 
\cite {HBech-WTittel} or four bases for a qutrit\cite{HBech-APeres}, 
but their security was only investigated against simple non-optimal attacks.

In this Letter, we investigate more general quantum cryptosystems where
the encoding is made into qudits with arbitrary $d$, extending 
on an earlier study by some of us that only considered simple 
individual attacks\cite{Bourennane}. A first protocol we study 
consists in using two mutually unbiased bases, just as in the original
BB84 scheme. The sender Alice 
sends a basis state in one of these two bases chosen at random, 
while the receiver Bob makes a measurement in one of these two bases, 
again at random. The basis
used by each party is subsequently disclosed on the public channel, 
so that Alice and Bob obtain correlated $d$-ary random variables 
if they used the same bases (and if there was no disturbance on the
channel), which happens with probability 1/2. 
The use of mutually unbiased (or complementary) bases implies that
if Alice and Bob use different bases, Bob's measurement
yields a random number that is uncorrelated with Alice's state.
The raw secret key is then made out of the correlated data (discarding the 
uncorrelated data is known as the sifting procedure).
This procedures ensures that any attempt by an eavesdropper Eve
(oblivious of the chosen basis) to gain information on Alice's state
induces errors in the transmission, which can then be detected by
the legitimate parties. The second qudit-based protocol that we study
makes use of all the $d+1$ mutually unbiased bases that are available in
a $d$-dimensional Hilbert space, much in the same way as the
six-state protocol for qubits. Here, Alice and Bob choose their basis
at random among the $d+1$ possible bases. This method clearly has a lower
yield than the first one since the sifting procedure only keeps 
one transmission out of $d+1$ (instead of 1/2). However, as we shall see,
this second protocol is more secure against individual attacks
in the sense that a slightly higher error rate is acceptable.

In what follows, we will analyze the security of these two cryptographic 
protocols against individual attacks (where the qudits are monitored
separately) as well as coherent attacks (where several
qudits are monitored jointly). For the individual case,
we consider a fairly general class of eavesdropping attacks that are 
based on quantum cloning machines. It is known for qubits that such
a cloning-based attack simply is the optimal eavesdropping
strategy, that is, the best Eve can do is to clone (imperfectly) Alice's
qubit and keep a copy while sending the original to Bob. An appropriate
measurement of the clone (and the ancilla system) after disclosure of the
basis enables Eve to gain the maximum possible information on Alice's 
key bit. Extending this cloning-based individual attack to higher dimensions
results in a lower bound on the information accessible to Eve 
for a given error rate. Hence, this yields an upper bound on the error rate,
which is a {\em necessary} condition for security against individual attacks.
(Higher error rates do not permit to establish a secret key
using one-way communication.)
We conjecture that applying the optimal cloner is actually
the best strategy for Eve in any dimension, so that
this bound is actually tight.
For the case of coherent attacks, we consider a situation where Eve
interacts with a qudit sequence of arbitrary (but finite) length,
and then uses the basis information to extract key information. 
In particular, we make use of an information-theoretic uncertainty principle
to derive a lower bound on Bob's information, or, equivalently,
an upper bound on the error rate. It is a {\em sufficient} condition
for the protocol to be guaranteed to generate a nonzero net key rate
for all attacks.

Let us consider first an individual eavesdropping based on
the use of a quantum cloning machine for qudits. 
The particular cloning machine that is best using depends on 
whether the protocol uses two bases or $d+1$ bases. We focus on
the case of two bases first, for which we need to use a cloner that
copies equally well two mutually unbiased bases, e.g. the computational
basis $\{\ket{k}\}$, with $k=0,\cdots,d-1$, and its dual under 
a Fourier transform
\beq
\overline{\ket{l}} 
= {1\over\sqrt{d}} \sum_{k=0}^{d-1} e^{2\pi i (k l/d)} \ket{k}
\eeq
with $l=0,\cdots,d-1$. We use a general class of cloning transformations
as defined in \cite{Cerf}. If Alice sends the state $\ket{\psi}$, 
the transformation reads
\beq  \label{nic1}
\ket{\psi}_A \to \sum_{m,n=0}^{d-1} a_{m,n} \; U_{m,n}\ket{\psi}_B
\ket{B_{m,-n}}_{E,E'}
\eeq
where $A$, $B$, $E$, and $E'$ stand for Alice's qudit, Bob's clone, 
Eve's clone and cloning machine, respectively. Here, the amplitudes $a_{m,n}$ 
(with $\sum_{m,n=0}^{d-1}|a_{m,n}|^2=1$) characterize the cloner, while
the states $\ket{B_{m,n}}_{EE'}$ are $d$-dimensional Bell states,
that is, a set of $d^2$ orthonormal maximally-entangled states
of two qudits,
\beq
|B_{m,n}\rangle_{EE'} = {1\over\sqrt{d}} \sum_{k=0}^{N-1}
{\rm e}^{2\pi i (k n / d)} \ket{k}_E \ket{k+m}_{E'}
\eeq
with $m,n=0,\cdots, d-1$. Note that the kets must be
taken modulo $d$ here. The operators $U_{m,n}$ defined as
\beq
U_{m,n} = \sum_{k=0}^{d-1} {\rm e}^{2\pi i (k n / d)}
\ket{k+m}\bra{k}
\eeq
form a group of qudit error operators, generalizing the Pauli matrices
for qubits: $m$ labels the shift errors (extending the bit flip $\sigma_x$)
while $n$ labels the phase errors (extending the phase flip $\sigma_z$).
Tracing the output joint state given by Eq.~(\ref{nic1})
over $EE'$ implies that Alice's state $\ket{\psi}_A$
is transformed, at Bob's station, into the mixture
\beq  \label{nic2}
\rho_B = \sum_{m,n=0}^{d-1} |a_{m,n}|^2 \;
U_{m,n}\ket{\psi}\bra{\psi}U_{m,n}^{\dagger}
\eeq
Thus, the state undergoes a $U_{m,n}$ error with probability $|a_{m,n}|^2$.
Note that $U_{0,0}=\openone$, implying that the state is left
unchanged with probability $|a_{0,0}|^2$. 
If Alice sends any state $\ket{k}$ in the computational basis,
the phase errors $(n\ne 0)$ clearly do not play
any role in the above mixture since $U_{m,n}\ket{k}=e^{2\pi i(kn/d)}\ket{k+m}$,
so Bob's fidelity can be expressed as
\beq
F = \langle k|\rho_B|k\rangle = \sum_{n=0}^{d-1} |a_{0,n}|^2
\eeq
In the complementary basis, we have $U_{m,n}\overline{\ket{l}}
=e^{-2\pi i (l+n)m/d}\, \overline{\ket{l+n}}$, 
so the shift errors ($m\ne 0$) do not play any role
and Bob's fidelity becomes
\beq
\overline{F} = \overline{\bra{l}} \rho_B \overline{\ket{l}} 
= \sum_{m=0}^{d-1} |a_{m,0}|^2
\eeq
For the cloner to copy equally well the states of both bases, we choose
a $d\times d$ amplitude matrix of the form
\beq   \label{matrix1}
a=\left( \begin{array} {cccc}
v & x & \cdots & x \\
x & y & \cdots & y \\
\vdots & \vdots & \ddots & \vdots \\
x & y & \cdots & y
\end{array}
\right)
\eeq
with $x$, $y$, and $v$ being real variables satisfying the
normalization condition $v^2+2(d-1)x^2+(d-1)^2 y^2=1$. Thus,
Bob's fidelity is $F=v^2+(d-1)x^2$ in both bases,
and the corresponding mutual information between Alice and Bob 
(if the latter measures his clone in the good basis) is given by
\beq  \label{Bobinfo}
I_{AB}= \log_2 d + F \log_2 F 
+ (1-F) \log_2 \left({1-F\over d-1}\right)
\eeq
since the $d-1$ possible errors are equiprobable.

Now, the clone kept by Eve can be shown to be in a state
given by an expression similar to Eq.~(\ref{nic2}) but with the amplitudes
$a_{m,n}$ replaced by
\beq
b_{m,n} = {1\over d} \sum_{m',n'=0}^{d-1} e^{2\pi i (nm'-mn')/d} \; a_{m',n'}
\eeq
that is, the Fourier transform of the $a_{m,n}$'s \cite{Cerf}. 
This corresponds to a matrix similar to Eq.~(\ref{matrix1}) but with
\beqa
x&\to& x'=[v+(d-2)x+(1-d)y]/d \nonumber\\
y&\to& y'=(v-2x+y)/d \nonumber \\
v&\to& v'=[v+2(d-1)x+(d-1)^2 y]/d
\eeqa
resulting in a cloning fidelity for Eve given by $F_E=v'^2+(d-1)x'^2$. 
Maximizing Eve's fidelity $F_E$ for a given value of Bob's fidelity $F$ 
(using the normalization relation) yields the optimal cloner:
\beq  \label{cloner2b}
x=\sqrt{{F(1-F)\over d-1}}, \quad
y={1-F \over d-1}, \quad 
v=F
\eeq
The corresponding optimal fidelity for Eve is
\beqa
\lefteqn{F_E = {F\over d}+{(d-1)(1-F) \over d} } \hspace{2.5cm} \nonumber \\
&& +{2\over d} \sqrt{(d-1)F(1-F)}
\eeqa
Let us see how Eve can maximize her information on Alice's
state. If Alice sends the state $\ket{k}$, then it is clear from 
Eq.~(\ref{nic1}) that Eve can obtain Bob's error $m$ simply 
by performing a partial Bell measurement (measuring only the $m$ index)
on $EE'$. Then, it appears from Eqs.~(\ref{matrix1}) and (\ref{cloner2b}) 
that, in order to infer Alice's state,
Eve must distinguish between $d$ non-orthogonal states (corresponding to all
possible values of $k$) with a same scalar product $(dF-1)/(d-1)$
for all pairs of states, regardless the measured value of $m$. 
Consequently, Eve's information $I_{AE}$ is simply given 
by the same expression as Eq. (\ref{Bobinfo}) 
but replacing $F$ by $F_E$. As a result, Bob's and Eve's information curves
intersect exactly where the fidelities coincide, that is, at
\beq
F=F_E= {1\over 2}\left(1+{1\over\sqrt{d}}\right)
\eeq

We now use a theorem due to Csisz\'{a}r and K\"{o}rner\cite{Csiszar}, 
which provides a lower bound on the secret key rate, that is, the rate $R$
at which Alice and Bob can generate secret key bits via privacy amplification:
if Alice, Bob and Eve share many independent realizations of a probability
distribution $p(a,b,e)$, then there exists a protocol that generates
a number of key bits per realization satisfying 
\beq  \label{csiszar}
R \geq \max(I_{AB}-I_{AE},I_{AB}-I_{BE})
\eeq
It is therefore sufficient that $I_{AB}>I_{AE}$ in order to establish
a secret key with a non-zero rate. If we restrict ourselves to one-way
communication on the classical channel, this actually is
also a necessary condition. Consequently, the quantum cryptographic
protocol above ceases to generate secret key bits precisely at the point where 
Eve's information attains Bob's information. In Table~\ref{table1}, 
we have computed the disturbance $D=1-F$ (or error rate) at which
$I_{AB}=I_{AE}$ (or $F=F_E$), that is, above which Alice and Bob 
cannot distill a secret key any more by use of one-way 
privacy amplification protocols. Strictly speaking, since we only conjectured
here that the cloning-based attack is optimal for all $d$, 
$D_2^{\rm ind}$ is actually an upper bound on $D$ 
that must necessarily be satisfied to generate secret key bits 
with one-way protocols.
(A tighter upper bound on $D$ might exist if the optimal individual attack
was not cloning, but we conjecture this is not the case).
Interestingly, we note that $D$ increases with the dimension $d$, 
suggesting that a cryptosystem based on qudits 
is more secure for large $d$.

Now, we consider the second protocol where all $d+1$ bases are used. The cloner
that must be used then is an asymmetric universal cloner\cite{Cerf}, 
characterized by an amplitude matrix of the same form as (\ref{matrix1}) 
but with $x=y$, 
the normalization relation becoming $v^2+(d^2-1)x^2=1$. Here, Bob's fidelity
is given by $F=v^2+(d-1)x^2=1-d(d-1)x^2$, so that the cloner is
characterized by 
\beq
x^2={1-F \over d(d-1)}, \qquad v^2={ (d+1)F-1 \over d}
\eeq
As before, Bob's information is given by Eq.~(\ref{Bobinfo}). Eve's clone
is characterized by a matrix of the same form, with
\beqa
x&\to& x'=(v-x)/d  \nonumber\\
v&\to& v'=[v+(d^2-1)x]/d
\eeqa
so the corresponding fidelity is $F_E=v'^2+(d-1)x'^2=1-d(d-1)x'^2$.
For deriving
Eve's information, we need first to rewrite the cloning transformation as
\beqa
\lefteqn{ \ket{k}_A \to  
{v-x \over \sqrt{d}} \; \ket{k}_B \sum_{l=0}^{d-1} \ket{l}_E \ket{l}_{E'}
} \hspace{1cm}\nonumber\\ &&
+ x \sqrt{d} \; \sum_{m=0}^{d-1} \ket{k+m}_B \ket{k}_E \ket{k+m}_{E'}
\eeqa
After the basis is disclosed, Eve's strategy is first to measure 
both $E$ and $E'$, the difference (modulo $d$) of the outcomes 
simply giving Bob's error $m$. Making use of $v-x=x'd$ and expressing
$x$ and $x'$ as functions of $F$ and $F_E$, it is easy to check that
the best Eve can do then is to use the state of her clone $E$ 
as an estimate of Alice's state $\ket{k}$. 
If Bob makes no error ($m=0$), which happens with probability $F$, 
then it yields the correct value of $k$ with probability
$(F+F_E-1)/F$, while it yields any other of the $d-1$ possibilities
$l\ne k$ with probability $(1-F_E)/[(d-1)F]$. In contrast, if Bob makes
an error ($m\ne 0$), then Eve obtains the right $k$ with probability one.
Consequently, the average mutual information between Alice and Eve 
conditionally on Bob's error $m$ can be written as
\beqa
\lefteqn{  I_{AE}= \log_2 d + (F+F_E-1) \log_2\left({F+F_E-1 \over F}\right)
} \hspace{2cm} \nonumber \\ 
&& + (1-F_E) \log_2\left({1-F_E \over (d-1)F }\right)
\eeqa
One can check that, for a given $F$, $I_{AE}$ is slightly lower here
than for the 2-bases protocol, which is consistent with the stronger
requirement that we put on the cloner.
Therefore, the fidelity $F$ at which $I_{AB}=
I_{AE}$ is slightly lower, and the corresponding disturbance $D=1-F$ is
slightly higher. In Table~\ref{table1}, we have shown 
the corresponding upper bound $D_{d+1}^{\rm ind}$ 
for several values of $d$, illustrating that there
is a slight advantage in using all $d+1$ bases, as for the
6-state protocol for qubits\cite{DBruss,HBech-NGisin}.

Our last result concerns the most general eavesdropping strategy which
consists in applying a coherent attack on a sequence 
of qudits of arbitrary (but finite) size $n$. Actually, our reasoning
is simpler to state with a single qudit, but it remains valid for
qudit sequences. We use an uncertainty principle due to Hall\cite{Hall}
that puts a limit on the sum of Bob's and Eve's information: if ${\hat B}$
and ${\hat E}$ are Bob's and Eve's observables applied on the qudit
sent by Alice, then
\beq
I_{AB}+I_{AE} \leq 2 \log_2\left( d \; \max_{i,j} |\langle b_i|e_j \rangle|
\right)
\eeq
where $\ket{b_i}$ and $\ket{e_j}$ are the eigenstates of ${\hat B}$ and
${\hat E}$, respectively. Since Eve has no way of guessing the basis used by
Alice, her optimal information is the same for the correct and incorrect bases.
Thus, one can bound $I_{AE}$ by assuming that Eve measures an observable
${\hat E}$ complementary to ${\hat B}$ (i.e. $|\langle b_i|e_j \rangle|
=d^{-1/2}, \forall i,j$) \cite{gisin}:
\beq
I_{AB}+I_{AE} \leq \log_2(d) 
\eeq
Using the discussion following Eq.~(\ref{csiszar}), we conclude that
$I_{AB} > \log_2(d)/2$ is a sufficient condition to warrant security
against coherent attacks if the key is made out of a large number 
of independent realizations of $n$-qudit sequences (i.e., if the key
is much longer than $n$).
Using Eq.~(\ref{Bobinfo}), this translates into a lower bound on $F$,
or, equivalently, an upper bound on $D$
\beq
F\log_2\left({1\over F}\right)+(1-F)\log_2\left({d-1\over 1-F}\right)
< {\log_2 d \over 2}
\eeq
which guarantees that one can distill secret key bits.
This bound for coherent attacks, shown in Table~\ref{table1}, 
exactly coincides with the well-known 11\% bound on the error
rate for $d=2$ due to Mayers\cite{Mayers}.

In summary, we have extended standard quantum cryptography to protocols 
where the key is carried by quantum states in a space of arbitrary
dimension $d$. We have used a general model of quantum cloning\cite{Cerf} 
in order to calculate the information accessible to an eavesdropper 
monitoring the qudits individually. This provides an upper bound 
on the error rate above which the legitimate parties cannot distill
a secret key by use of one-way privacy amplification protocols 
(since $I_{AB}<I_{AE}$). We conjectured that this bound is tight
(i.e., applying the optimal cloner is the best strategy for Eve
to gain the maximum information). Our analysis also suggested that
the 2-bases protocol should be preferred to a $(d+1)$-bases one
since its maximum acceptable error rate is only slightly lower, while
the corresponding secret key rate is much larger. Finally, 
we have derived a very simple security proof of quantum cryptography
with qudits that exploits an intuitive information inequality 
constraining the simultaneous measurement of conjugate observables\cite{Hall}.
This results in an upper bound on the acceptable error rate that is more
restrictive than the previous one, but guarantees that a non-zero
secret key rate can always be produced (even with coherent attacks 
on sequences of finite length).
In the region between these two bounds, it is unknown
whether the security is guaranteed or not. It should be stressed that
all the bounds on $D$ discussed above tend to 1/2 for $d\to\infty$,
reflecting the advantage of using higher-dimensional spaces. However, 
practical limitations might be more severe in realistic qudit-based 
cryptosystems, in particular the influence of the detector's 
quantum efficiency and dark count rate.
This is discussed in a related paper\cite{long-paper}.

{\it Note:} After completion of this work, it was proven in an
independent paper\cite{bruss-new} 
that the optimal individual attack for qutrits ($d=3$)
when using all four mutually unbiased bases exactly coincides 
with our results based on the optimal cloning machine, 
as conjectured here.

\begin{table}
\narrowtext
\begin{tabular}{clll}
$d$ & $D^{\rm ind}_2$ (\%) & $D^{\rm ind}_{d+1}$ (\%) & $D^{\rm coh}$ (\%) \\
\hline
2 & 14.64 & 15.64 & 11.00 \\
3 & 21.13 & 22.67 & 15.95 \\
4 & 25    & 26.66 & 18.93 \\
5 & 27.64 & 29.23 & 20.99 \\
10& 34.19 & 34.97 & 26.21
\end{tabular}
\caption{Disturbance $D=1-F$ (or error rate) as a function 
of the dimension $d$. The columns $D_2^{\rm ind}$ and $D_{d+1}^{\rm ind}$
display the values of $D$ at which $I_{AB}=I_{AE}$ for a cloning-based
individual attack with the 2-bases or $(d+1)$-bases protocol, respectively. 
The last column $D^{\rm coh}$ corresponds
to an upper bound on $D$ that guarantees security against coherent attacks.}
\label{table1}
\end{table}

\end{multicols}

\end{document}